\documentclass[twocolumn,
prd,secnumarabic,nobalancelastpage,
amsmath,amssymb,superscriptaddress,nofootinbib]
{revtex4-2}

\usepackage{graphicx} 
\usepackage{xcolor}
\usepackage{hyperref}

\newcommand{\req}[1]{Eq.\,(\ref{#1})} 
\newcommand{\rs}[1]{section~\ref{#1}} 
 
\newcommand{\rf}[1]{Fig.\,\ref{#1}}

\begin{document}

\title{Noise mitigation in quantum enhanced fiber optic gyroscopes}

\author{Stefan Evans}
\affiliation{Naval Information Warfare Center Pacific, San Diego, CA 92152, USA}
\author{Joanna Ptasinski}
\affiliation{Naval Information Warfare Center Pacific, San Diego, CA 92152, USA}

\begin{abstract}
We analyze noise in a quantum-enhanced fiber optic gyroscope (FOG), focusing on one of the leading sources of phase uncertainty - uncorrelated photon saturation. Taking a squeezed state input as a source for N00N states, we compute the uncorrelated false coincidence counts at the optimal phase bias, and determine an upper limit to the squeezed amplitude $\xi$ which allows for sub-shot noise precision. As examples, we apply parameters of present-day quantum FOG experiments, and determine the maximum possible precision enhancement based on their respective $\xi$ and optimal phase bias points. Aiming to future FOG setups with higher N00N state fluxes, our result highlights the need to transition to multimode states to bypass the $\xi$ limitation, such as photon pairs generated by the dynamical Casimir effect.
\end{abstract}


\maketitle

\section{Introduction}


The super-resolution of entangled photonic N00N states~\cite{Dowling_2008} offers an improved phase sensitivity over classical optical instruments. Originally posed as a photonic analogy to the de Broglie wavelength~\cite{Jacobson_1995}, in which the effective wavelength governing interference is reduced by the entangled photon order, the N00N state interference exhibits resolutions inaccessible from the classical lens. A notable example is the photon deposition resolution  exploited for lithography~\cite{Boto_2000}.

Our focus is the fiber optic gyroscope (FOG)~\cite{Culshaw_2006,Lefevre_2022}. Building upon  approaches to a quantum-enhanced Sagnac interferometer~\cite{Mehmet_2010, Restuccia_2019}, recent experiments applied polarization and path entangled $N=2$ photon order N00N states to FOGs, demonstrating the sought-after Sagnac phase super-resolution. The enhanced resolution permits sub-shot noise precision, which improves with the entangled photon number.

Sub shot noise precision means that quantum FOGs surpass their classical counterparts when operating at the same photon flux. However, there remains a sizable gap between the quantum and classical FOG fluxes (MHz vs. $10^{13}$/s). Among the obstacles to reaching higher flux $N=2$ N00N states is the saturation of the signal with uncorrelated photons. Alternatively, higher $N>2$ N00N state orders as high as $N=6$~\cite{Resch_2005, Afek_2010} have been accomplished, although here too there remains the complication of low fluxes in such post-selected combinations of quantum and classical light.

In this work we analyze the noise arising from uncorrelated photon saturation, extracting a nontrivial dependence on the squeezed amplitude. This allows us to improve the phase uncertainty estimation and refine the optimal domains of phase bias for which sub shot-noise precision is possible.

We give a brief overview of the N00N state FOG in~\rs{phaseideal}. In~\rs{phaseGeneral} we look at the phase uncertainty, and insert a squeezed state to determine false coincidence counts and the resulting upper limit to the squeezed amplitude in~\rs{SecUpperXi}. In~\rs{experimentTest} we apply the parameters of today's experimental quantum FOGs~\cite{Fink_2019,Silvestri_2024}. As an outlook, we discuss an upcoming quantum FOG experiment, along with prospects for higher order and flux photonic N00N states. 

\section{Sub shot noise resolution}
\label{phaseideal}

We recall the Sagnac-Laue phase phase acquired between two counter-propagating beams in a fiber coil~\cite{Lefevre_2022}
\begin{align}
\label{SLphase}
\phi_{\rm SL}
=\frac{4\pi \Omega L r}{\lambda c}
\;,
\end{align}
with $L$ labeling fiber length, $r$ the coil radius, and $\Omega$ the angular velocity. Classical FOG measurements of~\req{SLphase} are limited to (Poissonian) shot noise precision
\begin{align}
\label{SNlimit}
\Delta\phi_{\rm SN}= 
1/\sqrt{M}
\;,
\end{align}
where $M$ is the number of photons measured.

Recently, a polarization-entangled N00N state FOG experiment surpassed the shot noise limit by exploiting photon-number dependent phase resolution~\cite{Fink_2019}:
\begin{align}
\label{N00Nsuper}
\left|\psi_{\rm N00N}\right> = \frac1{\sqrt{2}}(\left|N0\right> - e^{iN(\phi_{\rm SL}+\phi_{\rm bias})}\left|0N\right>)
\;.
\end{align}
The photon order $N$ enhancement to the Sagnac phase sensitivity enters the coincidence count as~\cite{Fink_2019, Silvestri_2024}
\begin{align}
\label{Correlmeas}
M_{\rm cc}(\phi_{\rm SL}+\phi_{\rm bias})=
\frac{M_{\rm N00N}}{2}\Big(1+\cos(N(\phi_{\rm SL}+\phi_{\rm bias}))\Big)
\;,
\end{align}
where $M_{\rm cc}$ is the number of coincident counts and $M_{\rm N00N}$ is the number of N00N state pairs delivered to detectors within a time window to be specified below. Throughout this work, total photon number $M$ and coincidence counts $M_{\rm cc}$ and $M_{\rm N00N}$  label unitless quantities i.e. flux times measurement time.

The $N$ N00N state order gives an effective enhancement to the total photon number which reduces the shot noise phase uncertainty:
\begin{align}
\label{phiQM}
\Delta\phi_{\rm QM}= 
\Delta\phi_{\rm SN}/\sqrt{N}=
1/\sqrt{NM}
\;,
\end{align}
where \lq QM\rq\ labels the quantum enhanced shot noise resolution. \req{phiQM} possesses a factor $\sqrt{N}$ smaller phase uncertainty compared to the classical case~\req{SNlimit}.

Considering the angular velocity at which the phase uncertainty is equal to the Sagnac phase: $\Omega_{\rm min}=\Omega(\phi_{\rm SL}=\Delta\phi_{\rm QM})$ is
\begin{align}
\label{OmegaMin}
\Omega_{\rm min}=\frac{\lambda c}{4\pi L r\sqrt{NM}}
\;.
\end{align}
This sets an order of magnitude regime for the lower limits to angular rotation sensitivity. Taking for example a 1km fiber loop with 40cm radius, a wavelength of $\lambda=1550$nm, $M=10^6$ photons measured in a given time interval, and an $N=2$ order N00N state, we have $\Omega_{\rm min}=6.5\cdot10^{-5}$ rad/sec. It is in this low intensity regime that the N00N state enhancement makes a decisive difference - we recall e.g. earth's rotation rate of $7.29\cdot10^{-5}$rad/s. Transitioning to higher intensities however requires that we revisit noise sources - in the next section we look at uncorrelated photon signals.

\section{Uncorrelated photon noise}
\label{uncorr}

\subsection{Phase uncertainty}
\label{phaseGeneral}

We consider the uncertainty in a N00N state enhanced phase measurement. As an addition to the variance in the desired signal coincidence counts in~\req{Correlmeas}, we have the spurious coincidences due to uncorrelated photons which produce a Sagnac + bias phase dependent uncertainty. From hereon we write $\phi_{\rm SL}$ labeling implicitly the full phase including the bias $(\phi_{\rm SL}+\phi_{\rm bias})$.

We consider an experiment run time $t_{\rm meas}$ spanning many detection time windows ($\tau_{\rm detector}$). The number of false counts scales with the probability $P_{\rm cc-false}$ of a single spurious count within $\tau_{\rm detector}$:
\begin{align}
\label{Mccfalse}
M_{\rm cc-false}=P_{\rm cc-false}\frac{t_{\rm meas}}{\tau_{\rm detector}}
\;,
\end{align}
and similarly the signal N00N state pairs delivered 
\begin{align}
\label{MN00Neq}
M_{\rm N00N}=P_{\rm N00N}\frac{t_{\rm meas}}{\tau_{\rm detector}}
\;.
\end{align}
Combing the two, we have a spurious count which adds to to~\req{Correlmeas}: 
$ M_{\rm cc}(\phi_{\rm SL})\to M_{\rm cc}(\phi_{\rm SL})+M_{\rm cc-false} $. The false count contribution can be extracted from the dark fringes ($N\phi_{\rm SL}=\pm\pi,\pm 3\pi,\cdots$) of coincidence count fitting functions and modeled by a limited visibility~\cite{Okamoto_2008,Fink_2019, Silvestri_2024}.

The object of interest is the deviation in the number of photons delivered during $t_{\rm meas}$, which was shown in~\cite{Okamoto_2008} to produce a nontrivial optimization of the phase bias angle. The deviation combines the signal photons in~\req{Correlmeas} and false coincidences:
\begin{align}
\Delta M_{\rm cc}(\phi_{\rm SL})=
\sqrt{M_{\rm cc}(\phi_{\rm SL})+M_{\rm cc-false}}
\;.
\end{align}
$\Delta M_{\rm cc}$ can be recast as a phase uncertainty $\Delta\phi_{cc}$ accompanying the Sagnac-Laue phase in the coincidence count expression~\req{Correlmeas}: 
\begin{align}
\label{CCerror}
\Delta M_{\rm cc}(\phi_{\rm SL})
=&\;
M_{\rm cc}(\phi_{\rm SL}+\Delta\phi_{cc})-M_{\rm cc}(\phi_{\rm SL})
\;.
\end{align}
We solve for $\Delta\phi_{cc}$ exactly without expanding as is usually done to first order derivatives:
\begin{align}
\label{CCerrorPhase}
\Delta\phi_{cc\mp}=&\;
\pm\frac1N
{\rm acos}\Big(\frac{2\Delta M_{\rm cc}}{M_{\rm N00N}}+\cos(N\phi_{\rm SL})\Big)
+\frac{2\pi n}N
-\phi_{\rm SL}
\;,
\end{align}
where $n=0,\pm1,\pm2,\cdots$ corresponds to the range $N\phi_{\rm SL}=[(2n-1)\pi,(2n+1)\pi]$. Accounting for symmetry,  the $\Delta\phi_{cc+}$ solutions are positive  valued and valid within $N\phi_{\rm SL}=[(2n-1)\pi,2n\pi]$, while $\Delta\phi_{cc-}$ are negative and valid within $N\phi_{\rm SL}=[2n\pi,(2n+1)\pi]$. We also note that $\Delta\phi^{-2}_{\rm cc}$ gives the Fischer information~\cite{Okamoto_2008}.

We recall that $\Delta M_{\rm cc}$ describes the distribution of coincidence counts which one needs to average over. For completeness, we note that deviations in counts on the order of $\Delta M_{\rm cc}$ take on both signs. We thus consider absolute values $\Delta\phi_{cc}=|\Delta\phi_{cc\mp}|$ hereon, accounting for the  ${\rm sgn}(\Delta M_{\rm cc})$ dependence in $\Delta\phi_{cc}$ by using the sign which produces the larger noise contribution at the optimal bias point.

We evaluate phase uncertainty in the well-defined phase regions where $\Delta\phi_{cc}$ as defined in~\req{CCerrorPhase} is real: $|\frac{2\Delta M_{\rm cc}}{M_{\rm N00N}}+\cos(N\phi_{\rm SL})|<1$. Within the range $(0,\pi/N)$, this condition implies 
\begin{align} 
\phi_{\rm max}<\phi_{\rm SL}<\pi/N-\phi_{\rm min}
\;,
\end{align}
where $\phi_{\rm max}$ is the offset from the maxima ($N\phi_{\rm SL}= 2\pi n$) and $\phi_{\rm min}$ from the minima ($N\phi_{\rm SL}= \pi(2n+1)$) at which the phase uncertainty becomes well defined. For $\phi_{\rm min}\ll1$ and $\phi_{\rm max}\ll1$, the nested solutions reduce to $N\phi_{\rm min}\approx
\sqrt{(1/M_{\rm N00N})
\big(\sqrt{4M_{\rm cc-false}+1}+1\big)}$ and
$N\phi_{\rm max}=2/\sqrt{M_{\rm N00N}}$. The undefined regions repeat periodically at the maxima ($|N\phi_{\rm SL}-2\pi n|<\phi_{\rm max}$) and minima ($|N\phi_{\rm SL}-\pi(2n+1)|<\phi_{\rm min}$) of the coincidence count expression~\req{Correlmeas}, corresponding to the divergent uncertainties in the first order derivative expansion of~\req{CCerror} (\cite{Okamoto_2008}). With an exact solution however, a conveniently chosen fitting where the fringe visibility is slightly exaggerated can give a well-defined phase uncertainty - we do not consider this here since it comes at the cost of phase measurement precision.

The phase uncertainties near the divergent points are well above the shot noise limit. Near the minima, where uncorrelated counts $M_{\rm cc-false}$ provide the dominant contribution, $\Delta\phi_{cc}(\pi/N-\phi_{\rm min})
%
%
=
(\sqrt2-1)\phi_{\rm min}$. This is a substantially larger phase uncertainty than the shot noise~\req{SNlimit}: $\phi_{\rm min}/\phi_{\rm SN}
=\sqrt{2/N}(M_{\rm cc-false})^{1/4}$ (unless $N\gg 2$, a point we return to in Appendix~\ref{HighN00N}). We are thus interested in the uncertainty at the optimal phase bias between the singular points, which we now determine as a function of squeezed amplitude and experimental parameters.

\subsection{Squeezed state input}
\label{SecUpperXi}

Squeezed states are the prevalent source of N00N states in quantum sensing experiments. We focus here on $N=2$ N00N state interferometry, evaluating the noise due to random in phase false coincidence counts arising from $4$ and higher photon number states generated in the down conversion process. Extensions to $N>2$ order N00N states are discussed in Appendix~\ref{HighN00N}.

We consider for simplicity the squeezed vacuum state
\begin{align}
\label{sqvac}
\left|\xi\right>= &\;
e^{(\xi^*\hat a^2 - \xi \hat a ^{\dagger2})/2}
\left|0\right>
\\ \nonumber
=&\;
\sum_{m=0}^\infty 
\frac{\sqrt{(2m)!}}{2^{m}m!}
\frac{(-e^{i\arg(\xi)}\tanh|\xi|)^{m}}{\sqrt{\cosh|\xi|}}
\left|2m\right>
\;,
\end{align}
where the photon number probability
\begin{align}
\label{sqvacDist}
P_{2m}^{\xi}=
|\!\left<2m|\xi\right>\!|^2
=
\frac{(2m)!}{(2^{m}m!)^2}
\frac{(\tanh|\xi|)^{2m}}{\cosh|\xi|}
\;,
\end{align}
and the average photon number
\begin{align}
\label{sqvacavg}
\bar n=&\;
\left<\xi\right| \hat a^\dagger \hat a \left|\xi\right>
=\sinh^2|\xi|
\;.
\end{align}

Since today's SPDC sources produce 2-photon state fluxes on the order of $10^6$ pairs per second, we consider the regime where 2-photon states dominate the SPDC output. The resulting average photon number $\bar n\sim 10^{-4}$ ($|\xi|\sim 0.01$) over a typical single photon counter time resolution window of $\tau_{\rm detector}={\mathcal O}(100{\rm ps})$. Comparing the 4-photon and 2-photon probabilities,
\begin{align}
\label{P4P2}
\frac{P_4^{\xi}}{P_2^{\xi}}=
\frac{3}{4}\tanh^2|\xi|
\sim 7.5\cdot 10^{-5}
\;,
\end{align}
and similarly $P_6^{\xi}/P_4^{\xi}=(5/6)\tanh^2|\xi|$. Given that in general $P_{2m}^{\xi}\gg P_{2m+2}^{\xi}$ for this  perturbative in $\xi$ regime, we focus on $P_4^{\xi}$ and neglect $P_6^{\xi}$ and higher orders.

The $|4\rangle$ state produces predominantly spurious coincident counts with random phase at the detection. Due to loss, a negligible amount of 4-photon N00N states with dell-defined phase relations survive. The majority of loss occurs after the optical elements used to generate the N00N states e.g. in the Sagnac interferometer, producing uncorrelated 2 and 3-photon states responsible for false coincidence counts.

To estimate the false coincidence probability, we consider the 4-photon state undergoing loss via beamsplitting with the vacuum environment. The probability of a 2 or 3-photon byproduct
\begin{align}
\label{Pccavg}
P_{\rm cc-false}=
\Big(6T^2(1-T)^2 + 4T^3(1-T)\Big) P_4^{\xi}
\;,
\end{align}
where $T$ labels the transmission coefficient. The 2-photon contribution (first term in parenthesis)  dominates for a transmission of $T\sim0.1$ relevant to FOG setups. In comparison, the probability of the desired $N=2$ N00N state reaching the detector scales with transmission squared:
\begin{align}
\label{PN00Nform}
P_{\rm N00N}=T^2P_2^{\xi}
\;.
\end{align}
Applying the above probabilities to~\req{Mccfalse} and~\req{MN00Neq} gives the uncorrelated and signal coincidence counts.

Following the steps in~\rs{phaseGeneral}, we evaluate the phase uncertainty in~\req{CCerrorPhase} for different squeezed amplitudes, transmission coefficients, and measurement times. We find in~\rf{biasPlot} that the minimum uncertainties at the optimal phase bias points grow with the squeezed amplitude. Note that we consider transmission coefficients stemming from optical loss, and not single photon detector quantum efficiencies, which as assume are large ($>95\%$, e.g. as in superconducting nanowires).

%
\begin{figure}[h]
\center
\includegraphics[width=0.99\columnwidth]{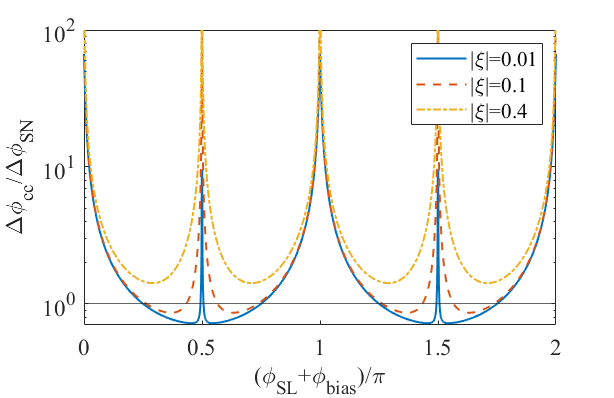}
\caption{Plot of phase uncertainty from~\req{CCerrorPhase} over shot noise~\req{SNlimit}, as a function of Sagnac-Laue and bias phase. For each squeezed amplitude shown, the detection timing window $\tau_{\rm detector}=100{\rm ps}$, transmission $T=0.1$ and count collection time $t_{\rm meas}=1000$ sec. 
\label{biasPlot}
} 
\end{figure}
%
%

We find an upper limit to squeezed amplitude for which sub shot noise precision is possible. To show this in more detail, we plot the minima in uncertainty as a function of $|\xi|$ in~\rf{xiLimit}. This limit grows with transmission ($|\xi|=0.181$ at $T=0.1$, and $|\xi|=0.406$ at $T=0.75$), yet is largely independent of the measurement time and detection time window. Note at larger transmissions we approach values of $|\xi|$ where higher photon number contributions ($>4$) to the false coincidence counts start to matter. These contributions further decrease the permissible $|\xi|$ range, making this result a conservative noise saturation estimate. A similar estimate can be derived for coherent state setups which rely on post-selection schemes, see Appendix~\ref{cohInput}.

%
\begin{figure}[h]
\center
\includegraphics[width=0.99\columnwidth]{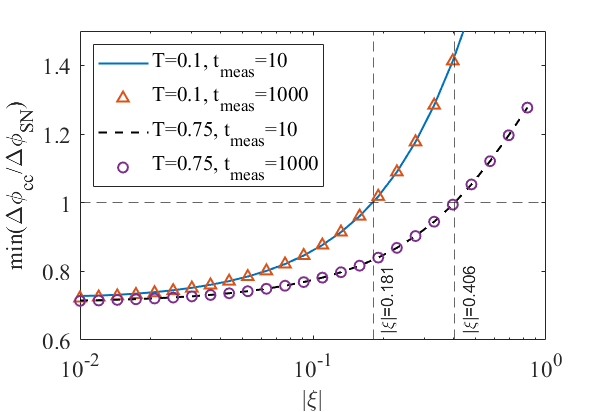}
\caption{Plot of spurious count phase uncertainty $\Delta\phi_{cc}$ over the shot noise as a function of squeezed amplitude. The detection time window $\tau_{\rm detector}=100{\rm ps}$ in each plot. 
\label{xiLimit}
} 
\end{figure}
%
%

\subsection{Experimental implications} 
\label{experimentTest}

We apply the above phase uncertainty considerations to recent quantum FOG experiments. We consider first the $N=2$ N00N state FOG setup in~\cite{Silvestri_2024}, the first path entanglement based FOG to resolve angular velocities below earth's rotation rate. This setup reported a transmission of $T=0.1$, $4000$ N00N state coincidence counts per second amounting to $M_{\rm N00N}=7.2\cdot10^6$ over $t_{\rm meas}=1800$ seconds, and a $\tau_{\rm detector}=156$ps detector timing resolution.

Following the steps in~\rs{uncorr}, these experimental parameters amount to a squeezed amplitude of $|\xi|= 0.011$. Applying the results of~\rf{biasPlot}, this is within the range of squeezed amplitudes for which sub shot noise phase uncertainty is possible. Additional dark photon counts beyond the higher photon number sourced flux considered above (e.g. sub kHz count per second thermal background) produce a negligible number to the false coincidence counts, and are easily mitigated by edgepass or bandpass filters.

While sub shot noise precision is possible for these experimental parameters, $\Delta\phi_{\rm cc}/\Delta\phi_{\rm SN}<1$ is only accessible within the (periodically repeating) range $\pi/4<\phi_{\rm SL}+\phi_{\rm bias}<0.493\pi$. At the optimal point $\phi_{\rm SL}+\phi_{\rm bias}=0.453\pi$, we have $\Delta\phi_{\rm cc}/\Delta\phi_{\rm SN}=0.723$, close to the ideal ($N=2$ N00N state enhanced) shot noise reduction factor of $1/\sqrt2=0.707$.

We recall that the experiment in~\cite{Silvestri_2024} measured a Sagnac phase of $5.5\pm0.4$mrad, an uncertainty larger than the $\Delta\phi_{\rm SN}=0.26$ shot noise. This measurement averages 11 data points taken at $\pi/8$ phase bias intervals, the majority of which located at points where uncertainty is larger than the shot noise according to~\rf{biasPlot}. The averaging of these measurements enlarge what may be an otherwise sub-shot noise resolution achieved at the optimal bias point.

We turn to an earlier experiment~\cite{Fink_2019} which pursued a similar $N=2$ N00N state FOG measurement. Here $t_{\rm meas}$=20ms data points were reported, each collecting~1956 photons, and a similar transmission of $T\sim 0.1$. We assume the same 156ps detection timing window  as in the above example.

These experimental parameters give $|\xi|=0.039$, also within the range of squeezed state amplitude for which sub-shot noise precision is possible. In this case, $\Delta\phi_{\rm cc}/\Delta\phi_{\rm SN}<1$ is accessible within $\pi/4<\phi_{\rm SL}+\phi_{\rm bias}<0.467\pi$. At the optimal point $\phi_{\rm SL}+\phi_{\rm bias}=0.403\pi$, we have $\Delta\phi_{\rm cc}/\Delta\phi_{\rm SN}=0.786$. Impressively, a factor 0.87 reduction of the shot noise was achieved in~\cite{Fink_2019} at the optimal phase bias point, close to the smallest achievable uncertainty ratio of 0.786 at this $|\xi|$.

\section{Conclusion}

We have computed the N00N state based phase measurement uncertainty as a function of an input squeezed state amplitude $\xi$. We found an upper limit to $|\xi|$ allowing for sub-shot noise precision. The crucial component is the uncorrelated photon flux as a function of  $\xi$ - without this contribution, an optimal phase bias point and achievable precision cannot be established theoretically, potentially leading one to overestimate the accessible quantum enhancement to the shot noise limit.

We applied this analysis to two recent $N=2$ N00N state FOG experiments. In one case~\cite{Fink_2019}, we determined the minimum phase uncertainty to be larger than the expected (ideal) factor $\sqrt2$  shot noise reduction. We found this corrected form to be close to the measured sub shot noise resolution achieved in~\cite{Fink_2019}. In the second experiment~\cite{Silvestri_2024}, a similar  bound to precision was obtained, but in this case the optimal phase bias point was not employed to achieve sub shot noise resolution.  Our result thus serves to refine the analysis of both prior and upcoming experimental measurements, allowing one to quantify the achievable sub shot noise precision and select the suitable bias angle to exploit it.

We recall that the analysis here is based on the assumption that all loss byproducts of the higher (than 2) photon number states have uncorrelated phase, since most loss occurs after the N00N states are generated. However, a fraction of loss still occurs in between the SPDC squeezed state source and the N00N state generation, in which case some residual phase bias dependence may remain - this will be revisited in follow up work.

Our analysis will be applied to an upcoming quantum FOG experiment detailed in~\cite{Evans_2025}, building on the work of~\cite{Fink_2019,Silvestri_2024}. The projected 6.65 dB loss ($T=0.216$, noting a typo in table 3 of~\cite{Evans_2025}) provides a competitive upper bound on $|\xi|$, allowing for sub shot noise precision angular velocity measurements below earth's rotation rate.

Looking towards future experiments, one will in principle need higher $N=2$ N00N state fluxes than the current bounds on $|\xi|$ permit for an SPDC source. A prime candidate for bypassing this limit is a multi-mode squeezed state~\cite{Grace_2020}. One may also consider the dynamical Casimir effect~\cite{Dalvit_2021a,Dalvit_2021b} - here the 4 and higher order photon states, which would otherwise be detrimental to the SPDC based setup, have different frequencies and thus can be easily filtered out.

\section*{Data availability}

Data sharing is not applicable to this article as no new data were created or analyzed in this study.

\acknowledgements
This work was supported by the Naval Information Warfare Center Pacific In-house Innovation Program.

\bibliographystyle{apsrev4-2}
\bibliography{refsNoise}

\appendix

\section{Coherent state input}
\label{cohInput}

The squeezed state analysis in~\rs{SecUpperXi} can also be applied to a coherent state - in this case N00N states are extracted via post selection~\cite{Kim_2021}. The average photon number $
\left<\alpha\right| \hat a^\dagger \hat a \left|\alpha\right>
=|\alpha|^2$ and the Poisson  distribution of photon number probabilities 
\begin{align}
\label{Pncohrnt}
P_n^\alpha=|\langle n|\alpha\rangle|^2 
= e^{-|\alpha|^2}\frac{|\alpha|^{2n}}{n!}
\;.
\end{align}

We consider $N=2$ N00N state phase measurements in the perturbative in $|\alpha|$ regime. Here the dominant contribution to uncorrelated coincidence counts is the 3-photon state which undergoes loss of one photon. The resulting false count probability 
\begin{align}
\label{PccavgCoh}
P_{\rm cc-false}=3T^2(1-T)P_3^\alpha
\;,
\end{align}
while the $N=2$ N00N state probability follows~\req{PN00Nform}.

We plot the resulting phase uncertainty~\req{CCerrorPhase} at optimal bias as a function of coherent state amplitude in~\rf{alpLimit}. A similar dependence on transmission follows as in the squeezed state case. We do not consider the added effect of 1-photon states which can saturate either of the two detectors, effectively reducing the quantum efficiency of the coincidence counting. This further reduces the upper bound to $|\alpha|$ seen in~\rf{alpLimit}. 

%
\begin{figure}[h]
\center
\includegraphics[width=0.99\columnwidth]{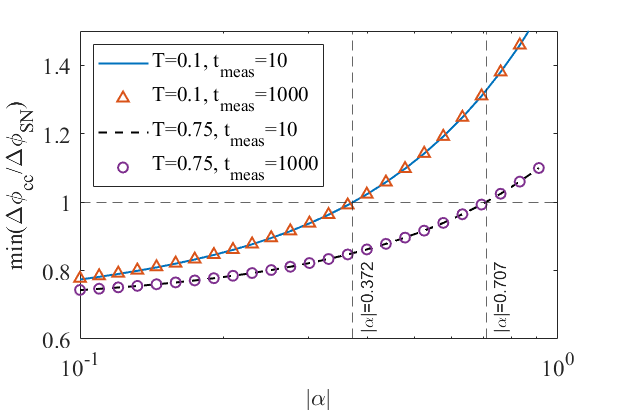}
\caption{Plot of spurious count phase uncertainty $\Delta\phi_{cc}$ over the shot noise as a function of coherent amplitude. As in~\rf{xiLimit}, the detection time window $\tau_{\rm detector}=100{\rm ps}$ in each case. 
\label{alpLimit}
} 
\end{figure}
%
%

\section{Higher N00N state orders}
\label{HighN00N}

Higher order ($N>2$) N00N state phase measurements have been performed using a combination of classical and squeezed light~\cite{Hofmann_2007,Afek_2010}. In this case one must optimize the combined squeezed and coherent state amplitudes.

We sum the probabilities of different coherent and squeezed state combinations giving a total of $N$ photons:
\begin{align}
P_N\equiv \sum_{m=0}^{\lfloor N/2 \rfloor}
P^{\alpha}_{N-2m}P^{\xi}_{2m}
\;,
\end{align}
using the coherent (\req{Pncohrnt}) and squeezed (\req{sqvacDist}) photon number probabilities.  Applying~\req{MN00Neq} and generalizing~\req{PN00Nform} to $N$ photons, the N00N state budget within $t_{\rm meas}$ is (up to a prefactor dependent on the beamsplitting)
\begin{align}
M_{\rm N00N}\approx 
P_N T^N\frac{t_{\rm meas}}{\tau_{\rm detector}}
\;.
\end{align}

As for the false coincidences in the perturbative $|\xi|$ and $|\alpha|$ regimes, we estimate the uncorrelated coincidence by the $P_{N+1}$ contribution after the loss of a single photon and $P_{N+2}$ after losing 2 photons:
\begin{align}
M_{\rm cc-false}\approx&\;
(N+1)T^{N}(1-T)\frac{t_{\rm meas}}{\tau_{\rm detector}}
\\ \nonumber
&\;\times
\Big(P_{N+1}+\frac N2P_{N+2}(1-T)\Big)
\;.
\end{align}
We plot the resulting phase uncertainty~\req{CCerrorPhase} for the $N=4$ and $N=5$ cases in~\rf{N4N5plot}. As expected, is it advantageous to have a larger contribution of photons by the coherent state.

%
\begin{figure}[h]
\center
\includegraphics[width=0.99\columnwidth]{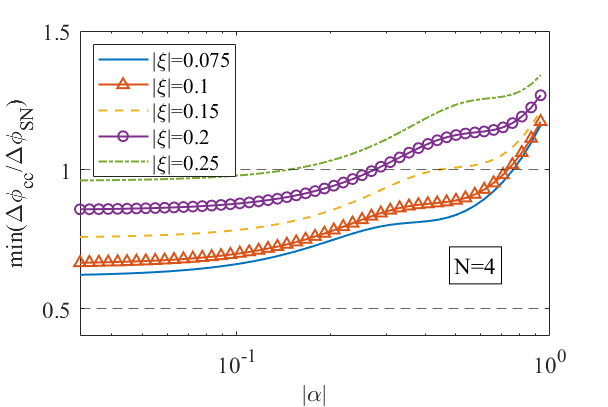}
\includegraphics[width=0.99\columnwidth]{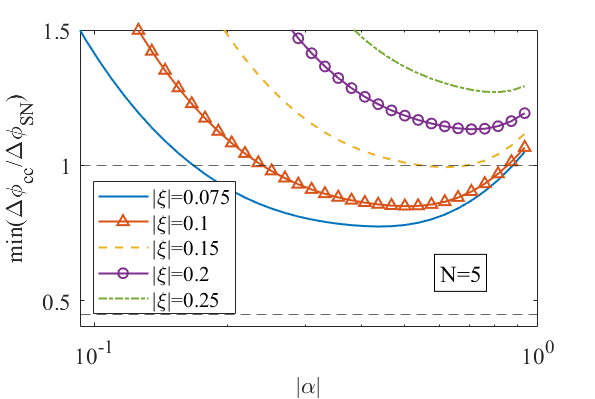}
\caption{Plot of $N=4$ (top) and $N=5$ (bottom) N00N state spurious count phase uncertainty $\Delta\phi_{cc}$ over the shot noise vs. coherent amplitude. Different fixed squeezed state amplitudes are plotted, and in each case $\tau_{\rm detector}=100{\rm ps}$, $T=0.3$ and $t_{\rm meas}=1000$. Dashed horizontal lines mark the boundaries of the sub shot noise regime (1, $1/\sqrt N$). 
\label{N4N5plot}
} 
\end{figure}

\section{Additional noise effects}

\subsection{Dispersion}
\label{AppCoherence}

We return to the expression for Sagnac output power~\req{Correlmeas} and introduce a finite linewidth: 
\begin{align}
\label{PmeasCoh}
M_{\rm cc}=\frac{M_{\rm N00N}}{2}\Big(1+C\cos(N\phi_{\rm SL})\Big)
\;,
\end{align}
which is accounted for in the coherence function~\cite{Lefevre_2022} 
\begin{align} 
\label{cohF}
C= e^{-3.5\Delta t^2/\tau^2}
\;,
\end{align}
where $\Delta t$ is the arrival time difference between the two light paths, and coherence time $\tau=1/\Delta \nu=\lambda^2/c\Delta\lambda$. A large linewidth does not compromise FOGs due to reciprocity (order fs arrival time differences $\Delta t\ll \tau$).

Non-reciprocal effects can also be minimized:  we recall that due to the polarization-dependent refractive index $n(\lambda)$, we have a difference in broadening between  polarizations along the fast and slow axes of the PM fiber: 
\begin{align}
\Delta t_{\rm chr}\equiv\Delta\tau_\parallel-\Delta\tau_\perp\neq0
\;,
\end{align} 
where $\Delta t_{\rm chr}$ labels the chromatic broadening - this is nonreciprocal as it grows with fiber length, where typical PM fibers exhibit $\Delta t_{\rm chr}/L\Delta\lambda\sim0.01$ps/km$\cdot$nm~\cite{Tang_2006}. This adds a factor to the coherence function in~\req{cohF}: $C\to Ce^{-3.5\Delta t_{\rm chr}^2/\tau^2}$, up to a constant in the exponential. Given a coherence time of $\tau\sim8$ps ($\Delta\lambda=1$nm, $\lambda=1550$nm), this is a very small reduction to the coherence function. Using a twisted fiber approach~\cite{Doerr_1994, Silvestri_2024} where the two input beams share the same fiber axis, this dispersion can be further reduced.

Turning to polarization, the broadening effect from polarization mode dispersion  (sub ps$/\sqrt{km}$) 
is also below the coherence time, leading to a small reduction of the coherence function~\cite{Chamoun_2015}, which can be combined with the spurious coincidence count analysis in the main text. We expect this effect to be sub-leading for the fiber lengths we consider; the absence of significant polarization non-reciprocities in PM fibers nearing 5km was demonstrated~\cite{Takei_2022} in a classical FOG. 
%

\subsection{Pump laser and SPDC source instabilities}

We briefly summarize some additional noise sources that are well documented for classical FOG experiments. We reiterate the prevalent effects which were adapted to the quantum FOG in~\cite{Evans_2025}. One source of noise is the wavelength instability in the pump field (distinguishing the center wavelength variance from the linewidth) driving the SPDC source. Another uncertainty comes from the intensity variance of the pump field. Combined, these two effects can produce large phase errors. However, they can be mitigated via signal processing of the pump beam and adjusting the predicted Sagnac phase shift.

In the case of nonlinear N00N state generation methods, the temperature dependence of the nonlinear crystal is also important. Considering a change in output SPDC wavelength of e.g. $\sim0.2{\rm nm}/^\circ$C, and a temperature control stability to $0.01^\circ$C, the resulting phase uncertainty is on the order of $10^{-6}$ of the Sagnac Laue phase. This accommodates signal to noise ratios nearing the classical domain i.e. the shot noise at $\sim 10^{12}$ photon counts per second.

\end{document}